# Impact of Event Encoding and Dissimilarity Measures on Traffic Crash Characterization Based on Sequence of Events

Yu Song[1], Madhav V. Chitturi[2], David A. Noyce[2]

1. University of Connecticut; 2. University of Wisconsin-Madison

**Abstract**

    Crash sequence analysis has been shown in prior studies to be useful for characterizing crashes and identifying safety countermeasures. Sequence analysis is highly domain-specific, but its various techniques have not been evaluated for adaptation to crash sequences. This paper evaluates the impact of encoding and dissimilarity measures on crash sequence analysis and clustering. Sequence data of interstate highway, single-vehicle crashes in the United States, from 2016-2018, were studied. Two encoding schemes and five optimal matching based dissimilarity measures were compared by evaluating the sequence clustering results. The five dissimilarity measures were categorized into two groups based on correlations between dissimilarity matrices. The optimal dissimilarity measure and encoding scheme were identified based on the agreements with a benchmark crash categorization. The transition-rate-based, localized optimal matching dissimilarity and consolidated encoding scheme had the highest agreement with the benchmark. Evaluation results indicate that the selection of dissimilarity measure and encoding scheme determines the results of sequence clustering and crash characterization. A dissimilarity measure that considers the relationships between events and domain context tends to perform well in crash sequence clustering. An encoding scheme that consolidates similar events naturally takes domain context into consideration.

# 1 Introduction

Characterization of traffic crashes has been emphasized by traffic safety researchers for studying crash patterns and identifying safety countermeasures, with the objectives of mitigating crash injury severity and ultimately preventing crashes (Abdel-Aty et al., 2005; Carrick et al., 2010; Klena II and Woodrooffe, 2011; Wu et al., 2016, 2018). Prior studies have stressed the importance of considering pre-crash information in crash characterization for better insights about crash progression (Snyder and Knoblauch, 1971; Cross and Fisher, 1977; Krull et al., 2000; Kusano and Gabler, 2013a, 2013b; Porter et al., 2018; Wu et al., 2016, 2018). The use of crash sequence data has been shown to be effective in developing a crash categorization that reflects the progression of crashes and correlates with crash injury severity outcomes (Wu et al., 2018, 2016). Crash sequences are sets of chronologically ordered pre-crash and crash events, which are usually extracted from police crash reports and are available in United States' national-level crash databases such as the Crash Report Sampling System (CRSS) and Fatality Analysis Reporting System (FARS) of the National Highway Traffic Safety Administration (NHTSA).

Large sets of crash sequences can be analyzed using sequence analysis methods adapted from fields such as bioinformatics and sociology (Wu et al., 2016). The fundamental task of sequence-based crash characterization is to compare and differentiate sequences, which involves a variety of measures and techniques. In biological and social sciences, sequence analysis methods have been developed, studied, and applied for more than 40 years (Needleman and Wunsch, 1970; Abbott, 1983; Studer and Ritschard, 2014; Cornwell, 2015). Domain knowledge is needed for sequence analysis because the definition and formation of sequences are domain-specific. Adaptation of sequence analysis to traffic crash study is still very recent, so there are no existing guidelines for processing sequence data or analyzing sequences.

The objective of this paper is to fill the gap in crash sequence analysis by investigating the impact of sequence encoding and dissimilarity measures on crash characterization based on sequence of events. Sequence analysis follows a procedure of data processing, sequence encoding, measuring sequence dissimilarity, sequence comparing, and clustering, as illustrated in Figure 1 (Cornwell, 2015; Wu et al., 2016; Song, 2021). The sequence analysis steps, especially encoding and comparison, are highly dependent on domain knowledge. However, few studies have discussed techniques of transferring crash report information into sequences, and none has compared different sequence encoding schemes for crash sequence analysis (Song et al., 2021). The core of sequence analysis is measuring sequence dissimilarity, which is also the basis of sequence clustering. Various sequence dissimilarity measures have been developed for application in biological, computer, and social sciences (Hamming, 1950; Levenshtein, 1965; Abbott, 1983; Kruskal, 1983; Studer and Ritschard, 2014; Cornwell, 2015). A comparison of dissimilarity measures is needed to adapt the most appropriate ones to crash sequence analysis, but no study has made such an effort yet.

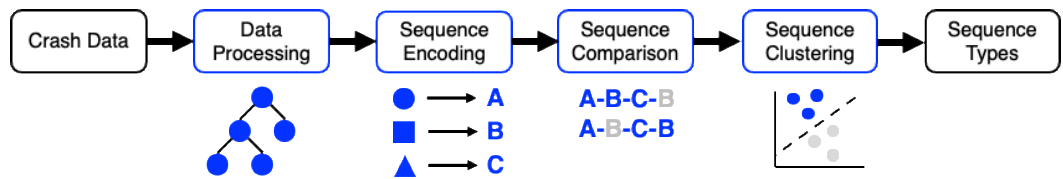

**Figure 1 Procedure of crash sequence analysis**

In this paper, sequences of single-vehicle crashes on interstate highways from the NHTSA CRSS database are analyzed, focusing on comparing the effects of different encoding schemes and dissimilarity measures on the crash sequence clustering results. Two types of encoding schemes, representing two levels of abstraction, and five types of dissimilarity measures, capturing different levels of context details, are applied. Clustering results are compared with a benchmark categorization of crashes provided by the CRSS database to evaluate clustering quality using different combinations of encoding schemes and dissimilarity measures.

Findings from this paper would be beneficial to traffic safety practitioners and researchers in identifying crash patterns and progressions, which can then be used to develop appropriate countermeasures and strategies to reduce crashes.

## 2 Literature Review

Sequence analysis methods such as sequence alignment were developed and applied for purposes such as protein or DNA sequence characterization, codes and error control, text and speech processing, and the study of social phenomena (Abbott, 1983; Kruskal, 1983). Early foundations of sequence comparison methods were set by researchers such as Hamming and Levenshtein who proposed measures of sequence dissimilarity, as well as Needleman and Wunsch who developed the optimal matching (OM) algorithm to efficiently quantify sequence dissimilarity (Hamming, 1950; Levenshtein, 1965; Needleman and Wunsch, 1970). Promoted by Abbot in the 1980s, sequence analysis based on the OM methods started to gain popularity in the sociology field (Abbott, 1995, 1983; Abbott and Tsay, 2000). Ever since, variants of OM and new sequence analysis methods have been developed to address various data conditions and study needs (Elzinga, 2003, 2005; Yujian and Bo, 2007; Hollister, 2009; Gauthier et al., 2010; Elzinga and Studer, 2015). Sequence dissimilarity measures have been compared, using real or simulated data, for applications in biological and social science research (Blackburne and Whelan, 2012; Robette and Bry, 2012; Studer and Ritschard, 2014, 2016; Kang et al., 2020).

Kun-feng Wu and colleagues were the first to apply sequence analysis in a traffic crash study (Wu et al., 2016, 2018). In their study to identify the relationship between crash sequences and crash injury outcomes, Wu et al. introduced the basics of sequence analysis to the traffic safety community, and applied OM and clustering to characterize a set of fatal single-vehicle run-off-road crash sequences from the NHTSA FARS database (Wu et al., 2016). The characterization of crash sequences was validated by estimating the agreement between the sequence clustering results and a benchmark crash categorization provided by FARS. The sequence clustering results were then used as a variable in crash injury severity estimation, which showed significant correlations between sequence-derived crash types and crash injury outcomes. Wu et al.'s study confirmed that informative and meaningful crash characterization could be derived from crash sequence analysis, and safety countermeasures could be identified to target crashes and injuries from a crash sequence perspective. In another study, Wu et al. built a risk matrix based on motorcycle crash sequences and identified sequences leading to the highest injury risks (Wu et al., 2018).

Sequence analysis has been applied to investigate patterns in automated vehicle (AV) crashes (Song et al., 2021). Sequences of events were extracted from the original crash reports and were encoded using self-developed encoding scheme and procedure. Subsequence frequencies and event transition rates were analyzed to explore subsequence-level patterns. OM method was applied to characterize the AV crash sequences into seven types. Based on the associations of sequence types with environmental conditions and crash outcomes, a scenario-based AV safety evaluation framework was proposed to embed crash-sequence-derived test scenarios. Through investigating AV crash sequences, a preliminary procedure for crash sequence analysis was proposed. However, more efforts are needed to develop a comprehensive methodology for crash sequence research. As Wu et al. pointed out, further investigations are needed for crash sequence encoding, dissimilarity measuring, and clustering techniques (Wu et al., 2016).

## 3 Data and Methodology

To examine how encoding and dissimilarity measures affect the characterization of crash sequences, five OM-based dissimilarity measures were selected and applied to a set of interstate highway single-vehicle crash sequences, encoded with two levels of abstraction. The crash sequence data was obtained from the 2016-2018 NHTSA CRSS database. Sequence clustering results were evaluated using several quality measures and were compared with a benchmark, the CRSS-provided crash typology.

Figure 2 visualizes our study design. First, crash sequences are encoded by two encoding schemes with different levels of abstraction and compared using five different dissimilarity measures. The process generates 10 dissimilarity matrices, which are then fed to Mantel test to evaluate correlations between dissimilarity measures, and to the clustering process to characterize sequences as different types. The clustering results are each compared with the benchmark crash categorization to evaluate the degree of agreement. Results from the correlation and degree of agreement analyses are the basis for evaluating the sequence modeling method and impact of encoding and dissimilarity measures.

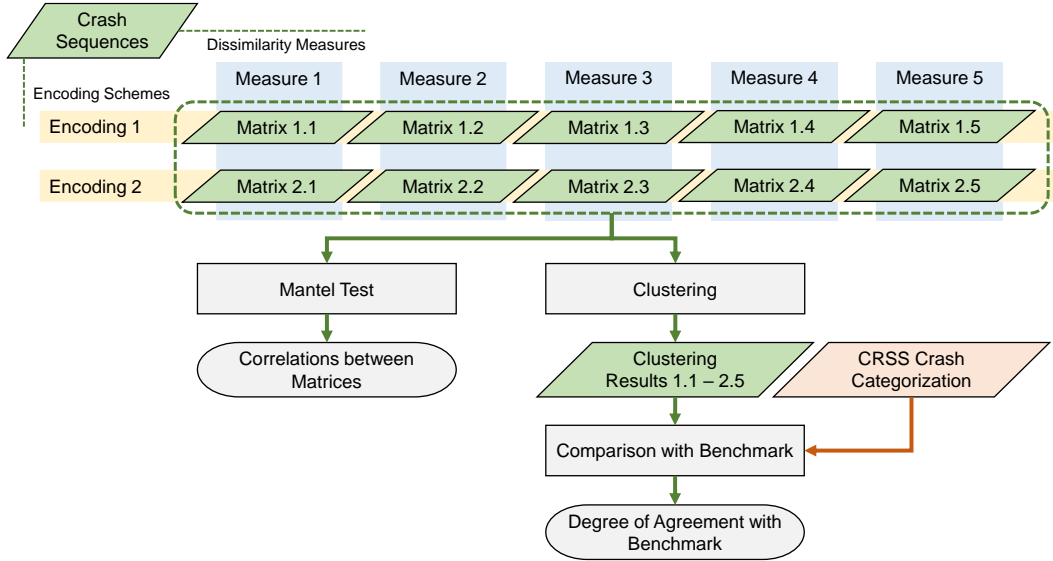

**Figure 2  Study design**

## 3.1  Crash Sequence Data

Single-vehicle crashes on interstate highways were obtained from the 2016-2018 CRSS crash data, using conditions (variables and values) listed in Table 1. The resulting data set consists of 2,676 observations, representing a weighted total of 385,484 crashes during the 2016-2018 period.

**Table 1  Conditions for obtaining data for case study from CRSS**

| Variable (Data Level) | Value | Description of Condition |
| --- | --- | --- |
| VE_FORMS (Crash) | = 1 | Only one vehicle-in-transport involved in crash. |
| INT_HWY (Crash) | = 1 | Crash occurred on interstate highway. |
| PVH_INVL (Crash) | = 0 | No parked/working vehicles involved. |
| WRK_ZONE (Crash) | = 0 | No work zone at crash location. |
| ALCHL_IM (Crash) | ≠ 1 | No alcohol-related crash. |
| BDYTYP_IM (Vehicle) | < 50 | Only automobile, utility vehicles or light trucks* involved. |
| TOW_VEH (Vehicle) | = 0 | No vehicle trailing involved. |
| BUS_USE (Vehicle) | = 0 | No bus involved. |
| SPEC_USE (Vehicle) | = 0 | No special use vehicles involved. |
| EMER_USE (Vehicle) | = 0 | No emergency use vehicles involved. |

Note: * Light trucks with Gross Vehicle Weight Rating (GVWR) ≤ 10,000 LBS

CRSS provides pre-crash event data in the VEHICLE data file via variables PCRASH1 (pre-event movement), PCRASH2 (critical event pre-crash), and PCRASH3 (attempted avoidance maneuver). The CEVENT data file provides a series of harmful and non-harmful events that occurred in the crashes, through the SOE (sequence of events) variable, ordered chronologically. Sequences analyzed in this case study were formed by combining the PCRASH1~3 variables and the SOE variable. Based on the CRSS definitions, PCRASH1~3 described "what a vehicle was doing just prior to the critical precrash event", "what made the vehicle's situation critical", and "what was the corrective action made, if any, to this critical situation", happened before the vehicle's SOE events. Single-vehicle crashes only had one vehicle-in-transport involved, therefore, the structure of such sequences was formed following the chronologically order of events as:

(PCRASH1 event) – (PCRASH2 event) – (PCRASH3 event) – (SOE events)

Lengths (number of elements) of the 2,676 crash sequences ranged from 4 to 12, as shown in Figure 3. The average length was 5.6.

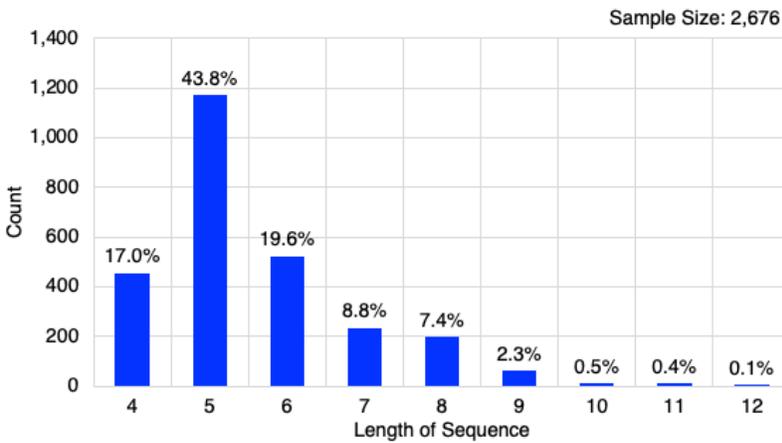

**Figure 3  Sequence lengths**

*3.2    Sequence Encoding*

Sequence encoding is a process of information abstraction and consolidation. The CRSS has converted crash information from the crash report to sequence of events following a standard reviewing and information extraction procedure (NHTSA, 2018). However, the original sequence encoding scheme provided by crash databases may need to be further consolidated to keep the number of distinct events manageable and the level of detail appropriate for specific analysis purposes (Wu et al., 2016).

In the CRSS interstate single-vehicle crash sequences, there are 123 distinct event types that make up the original encoding scheme. A new encoding scheme was developed to consolidate the original scheme by combining event types that were similar in nature. The process of developing the new encoding scheme is illustrated in Figure 4. The new scheme consists of 59 event types. From the original to the new scheme, the level of detail decreases, and the level of abstraction increases. As a result of applying different sequence encoding schemes on the interstate single-vehicle crash data, the original scheme formed 1,535 distinct sequences and the new scheme formed 1,105 distinct sequences. Details of the encoding schemes are presented in Table A - 1 in Appendix A.

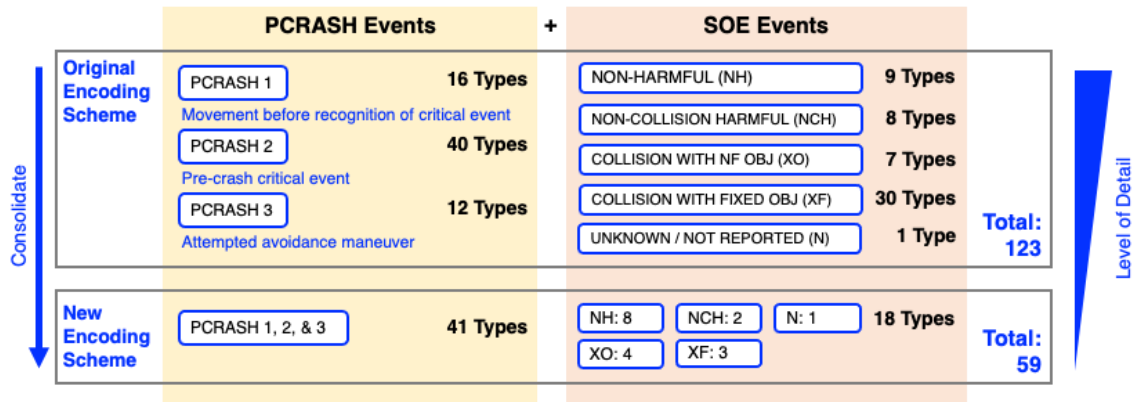

**Figure 4** Process of developing crash sequence encoding

Table 2 provides an example to show the logic of event consolidation. Roadside fixed object collision events were encoded in the three schemes, providing different levels of details about the collision events. In the original encodings, there were 30 types of fixed object collisions ("XF"). In the new encoding scheme, the 30 fixed object collision types were grouped into three categories based on their propensities toward injury risks (ranked "XFA", "XFB", and "XFC", from low to high, based on object hardness and potential contact area).

**Table 2** Example of encoding schemes

| Original | Scheme 1 | Scheme 2 | CRSS Description |
|---|---|---|---|
| 1v23 | XFB | XF | 23 Bridge Rail (Includes Parapet) |
| 1v24 | XFA | XF | 24 Guardrail Face |
| 1v25 | XFB | XF | 25 Concrete Traffic Barrier |
| 1v26 | XFB | XF | 26 Other Traffic Barrier |
| 1v30 | XFC | XF | 30 Utility Pole/Light Support |

*3.3   Dissimilarity Measures*

There are many ways to quantify sequence dissimilarity, which are the basis of sequence comparison and clustering (Studer and Ritschard, 2014, 2016). Intuitively, for the comparison of crash sequences, the most important attributes determining dissimilarity are the distinct elements and the order of elements (i.e., what happened and in what sequence did they happen). Depending on the encoding, events sometimes repeat in crash sequences (e.g., hitting multiple fixed objects), so event frequency is another potential attribute contributing to the dissimilarity between crash sequences. The sensitivity of OM-based dissimilarity measures to distinct elements, order of elements, and frequency of elements are higher than other types of dissimilarity measures, based on prior studies (Robette and Bry, 2012; Studer and Ritschard, 2014, 2016). Therefore, OM-based dissimilarity measures were selected for this study.

OM-based dissimilarity measures define several operations that can be used to transform one sequence to another (the procedure is called "alignment"), with costs assigned to the operations. The dissimilarity between the two sequences is then calculated as the smallest transformation cost needed (Abbott, 1983, 1995; Kruskal, 1983; Studer and Ritschard, 2014, 2016; Cornwell, 2015). Multiple types of operations can be applied in OM, including substitutions, deletions and insertions (or indels), compression and expansions, and transpositions (or swaps) (Kruskal, 1983; Studer and Ritschard, 2014, 2016). Substitutions and indels are the two commonly used operations.

A formal mathematical expression of OM dissimilarity between a pair of sequences, $x$ and $y$, is (Yujian and Bo, 2007; Studer and Ritschard, 2014, 2016):

$$d_{OM}(x,y) = \min_j \sum_{i=i}^{\ell_j} \gamma(T_i^j) \tag{1}$$

where $\ell_j$ denotes the transformations needed to turn sequence $x$ into $y$; $\gamma(T_i^j)$ is the cost of each elementary transformation $T_i^j$ (e.g., indel or substitution).

Here we use an example of two sequences "ABCD" and "ACB" to illustrate the principles of transformation operation costs and OM. Table 3 shows two ways to align the two sequences, with different combinations of operations. An indel costs d and a substitution costs s. Alignment 1 used 1 deletion and 2 insertions, costing 3d. Alignment 2 used 1 deletion and 1 substitution, costing d + s. There are many other ways to align "ABCD" and "ACB", with different costs. The OM method applies the Needleman-Wunsch algorithm and returns the minimum alignment cost as the dissimilarity between the two sequences (Needleman and Wunsch, 1970).

**Table 3 Sequence alignment costs**

| Sequence 1 | A | B | C | D | |
|---|---|---|---|---|---|
| Sequence 2 | A | C | B | | |
| **Alignment 1** | | | | | |
| Sequence 1 | A | | B | C | D |
| Sequence 2 | A | C/ | B | ø | ø |
| Cost | | d | | d | d | = 3d |
| **Alignment 2** | | | | | |
| Sequence 1 | A | B | C | D | |
| Sequence 2 | A | ø | C | <u>B</u> | |
| Cost | | d | | s | = d+s |

Note: Insertion is marked with ø,
Deletion is marked with slash/,
Substitution is marked with <u>underline</u>

The indel and substitution costs can be set by researchers based on specific problems and data. The cost schemes reflect how the differences between sequence elements are measured, and what real-world meanings are conveyed by the indel and substitution operations. According to prior studies, the cost schemes can be basic, with fixed indel and substitution costs; data-driven, with substitution costs determined by data; or localized, with indel costs based on sequence contexts.

3.3.1 Basic Optimal Matching Costs

Basic OM cost scheme assigns constant costs to indels and substitutions. Measures such as Levenshtein, Levenshtein II, and Hamming use constant costs but allow operations differently (Hamming, 1950; Levenshtein, 1965; Cornwell, 2015). For crash sequences, indels are a representation of addition or removal of certain actions/events/objects, and substitutions represent replacement of certain actions/events/objects with others. Under this general principle of OM for crash sequences, the choice of indel and substitution costs should reflect real-world meanings and difficulties of conducting operations of additions, removals, and replacements. However, the basic Levenshtein distance uses constant substitution cost of 2 and indel cost of 1, treating different replacements equally, meaning a replacement (A → B) costs the same as (A → C).

Past researchers have criticized OM on the lack of real-world meanings for the transformation operations(Wu, 2000; Levine, 2000; Abbott and Tsay, 2000; Abbott, 2000; Studer and Ritschard, 2014, 2016; Cornwell, 2015). To address this issue, more recent research has developed methods to define substitution and indel costs that sophisticatedly consider the real-world contexts of sequences (Hollister, 2009; Studer and Ritschard, 2014, 2016).

### 3.3.2 Data-driven Optimal Matching Costs

A commonly used data-driven method is to set up substitution cost based on the element transition rates (TRATE) observed from the sequence data set. The transition rate between two elements, A and B, is calculated as the probability of element A followed by element B in all cases that element A appears in the observed element space (Studer and Ritschard, 2014; Cornwell, 2015):

$$p(AB) = p(B_{p+1}|A_p) = \frac{n(AB)}{n(A)} \quad [2]$$

where *n(AB)* is the count of times that AB appears at consecutive positions (*p* and *p+1*) following that order; and *n(A)* is the count of times A appears in the element space. The TRATE-based symmetrical substitution cost is calculated as (Studer and Ritschard, 2014):

$$\gamma_{tr}(A,B) = 2 - P(AB) - P(BA) \quad [3]$$

By subtracting the transition rates from the basic Levenshtein substitution cost of 2, the TRATE-based substitution cost considers easier substitutions between elements that appear adjacent in observed sequences, compared with those that do not appear adjacent.

Although commonly used, the TRATE-based substitution cost is considered questionable due to equating high transition to high similarity (Studer and Ritschard, 2014). A more elaborate substitution cost was proposed to address this issue by considering the probability of events sharing a common future (Rousset et al., 2012; Studer and Ritschard, 2014). The idea of shared-future based substitution cost is that if the data shows a high probability of two events, A and B, sharing a same event, C, over a (user defined) position lag of q, then the A and B is considered highly similar and the substitution cost between A and B should be low. Mathematical expression of such a substitution cost is:

$$\gamma_{sf}(A,B) = \sum_{C \in E} \frac{\left(p(C_{+q}|A) - p(C_{+q}|B)\right)^2}{\sum_{F \in E} p(C_{+q}|F)} \quad [4]$$

where E is the space of all possible elements, and $p(C_{+q}|F)$ is the probability of F followed by C over q positions. For crash sequences, this data-driven substitution cost based on shared future is more suitable compared with the TRATE-based substitution cost.

### 3.3.3 Localized Optimal Matching

Apart from the substitution cost, the indel cost can also be more realistic and incorporate context better. A Localized OM (LOM) calculates indel costs based on whether the element being inserted or deleted is the same with its adjacent elements (Hollister, 2009; Studer and Ritschard, 2014). An indel of an element different (or more dissimilar) to its adjacent elements is considered more difficult and should be of a higher cost. In LOM, the indel cost of inserting an element U between elements A and B is calculated as:

$$c_I(U|A,B) = e\gamma_{\max} + g\frac{\gamma(A,U) + \gamma(B,U)}{2} \quad [5]$$

where *γ* indicates substitution cost. Thus *γ*$_{max}$ is the maximum substitution cost, *γ(A,U)* is the substitution cost between A and U, and *γ(B,U)* is that between B and U. Parameters *e* and *g* control the weights assigned to two components of the function, 1) the maximum substitution cost and 2) the average of distances between the element to be inserted and its neighbors. The first component is a global cost for conducting an indel, and the second component is a local cost considering the difficulty of transition from adjacent event to the inserted one (or vice versa). The *e* and *g* parameters are set by researchers based on their domain knowledge and research needs. One of the goals of using LOM is to avoid using a pair of indels in lieu of a substitution (violating the triangle inequality) (Hollister, 2009). Therefore, the values of e and g should satisfy $2e + g \geq 1$.

### 3.3.4 Cost Scheme of Selected Dissimilarity Measures

Cost scheme of the five dissimilarity measures is summarized in Table 4. In this paper, the Levenshtein distance, two types of data-driven substitution costs, and two types of localized indel costs were considered in setting up the cost schemes. Cost schemes of the two LOM measures are determined by parameters *e* and *g*, for which a range of values were tested in a sensitivity analysis and the values yielding the best clustering performance were selected. Details of the LOM parameter sensitivity analysis are covered in the Results and Discussion section.

**Table 4 Dissimilarity measures**

| Dissimilarity Measure | | | Cost Scheme |
|---|---|---|---|
| Optimal Matching | Levenshtein Distance | "OMlev" | Levenshtein (substitution cost = 2, indel = 1) |
| | TRATE-based OM | "OMtr" | TRATE-based substitution cost, indel = 1 |
| | Shared-future-based OM | "OMsf" | Shared-future-based substitution cost, indel = 1 |
| Localized Optimal Matching | TRATE-based Localized OM | "LOMtr" | TRATE-based substitution cost, indel cost parameters: e = 0.0~0.4, g = 0.8~0.2 |
| | Shared-future-based OM | "LOMsf" | Shared-future-based substitution cost, indel cost parameters: e = 0.0~0.4, g = 0.8~0.2 |

*3.4 Clustering*

Commonly used clustering methods such as hierarchical clustering and k-medoids clustering are applicable to crash sequence clustering. Whichever method is selected, clustering determines how sequences are grouped based on the sequence dissimilarity matrix. Sequence dissimilarity matrix consists of dissimilarity values (or distances) between all pairs of sequences, calculated using a selected dissimilarity measure. The k-medoids clustering method was selected for this paper because it is considered to be good at handling categorical data and outliers (Jin and Han, 2017; Nitsche et al., 2017). To compare the performance of dissimilarity measures in clustering, identical clustering settings were used on all dissimilarity matrices, and the clustering results were compared with the same benchmark crash categorization.

The CRSS crash types (derived from the variable "ACC_TYPE") was used as a benchmark crash categorization in this paper. As defined by CRSS, the ACC_TYPE variable "identifies the attribute that best describes the type of crash this vehicle was involved in based on the first harmful event and the precrash circumstances". Although crash sequences conveyed more information than the first harmful event and precrash circumstances, the ACC_TYPE categorization was considered feasible as a benchmark for dissimilarity measure comparison because this categorization could partially reflect the sequence patterns.

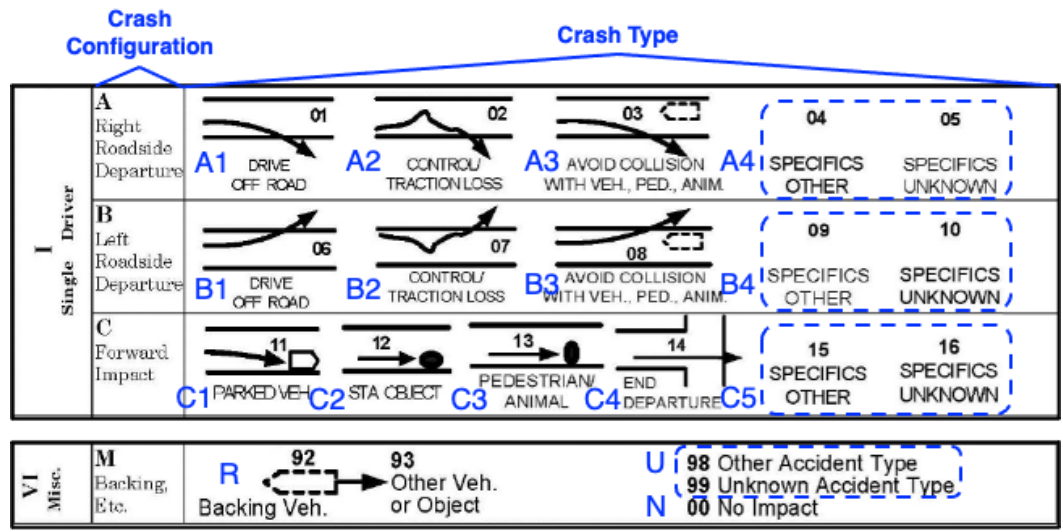

Note: Type C1 did not occur in the studied sample of interstate single-vehicle crashes.
**Figure 5  CRSS configurations and types of single-vehicle crashes**

As shown in Figure 5, CRSS categorized interstate single-vehicle crashes into roughly 15 types (C1 is not in the studied sample). To match the number of crash types in the benchmark categorization, the target cluster number, k, was also set to 15 for the weighted k-medoids clustering. To confirm that k=15 was an appropriate value, for each combination of encoding scheme and dissimilarity measure, a range of k values were tested with clustering quality indices plotted. An example of clustering quality measures for the OMlev measure under the two encoding schemes are shown in Figure 6.

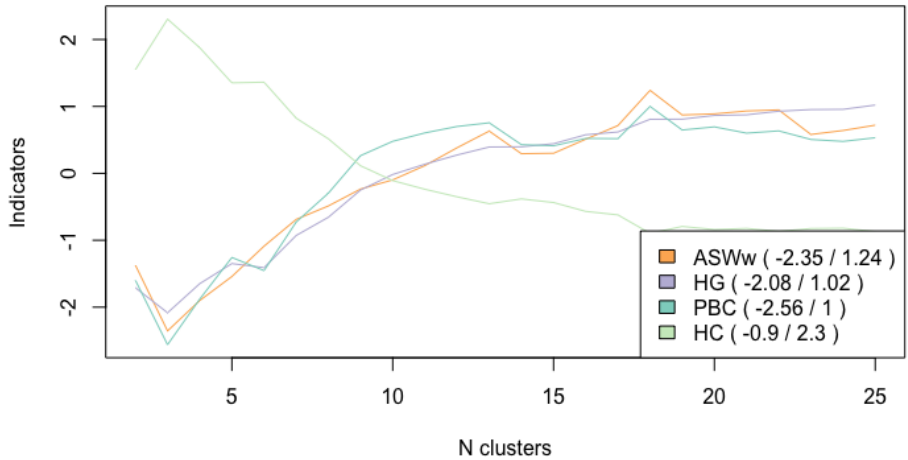

**(a)  Original Encoding**

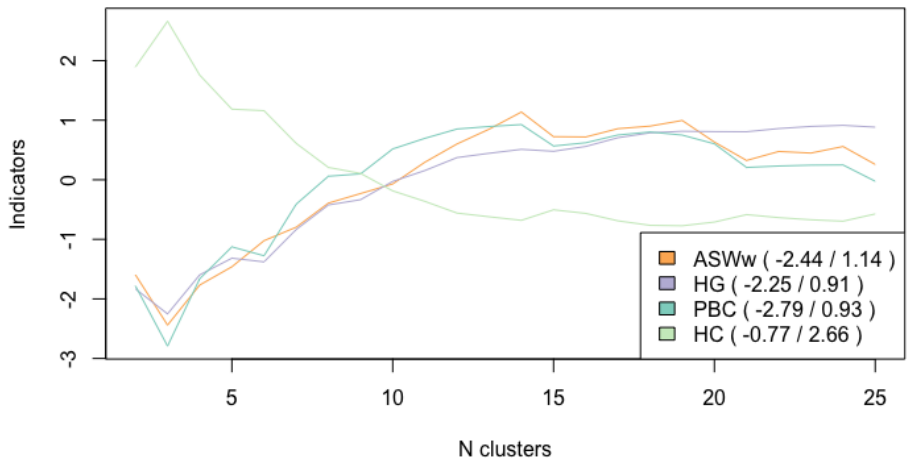

**(b)  New Encoding**

**Figure 6  Clustering quality of the OMlev measure**

Four clustering quality indices were employed for evaluation, the Weighted Average Silhouette Width (ASWw), Hubert's Gamma (HG), Point Biserial Correlation (PBC), and Hubert's C (HC) (Studer, 2013). An optimal k value would generate maximum ASWw, HG, and PBC (all range from -1 to 1), and minimum HC (ranges from 0 to 1). The charts in Figure 6 present standardized values of clustering quality indices for easier identification of good k values, and k=15 was shown to ensure satisfactory clustering quality. When k=15, the ASWw, HG, and PBC values were considerably large, and the HC value was considerably small.

## 3.5 Result Evaluation

### 3.5.1 Correlation

Mantel test is a correlation test for distance matrices, first developed by Nathan Mantel (Guillot and Rousset, 2013; Kang et al., 2020; Mantel, 1967; Robette and Bry, 2012). Using each dissimilarity measure, we obtained an n-by-n dissimilarity matrix of crash sequences, with n equals to the crash sample size of 2,676. Each dissimilarity matrix had n(n – 1)/2 elements. Five dissimilarity matrices were obtained from applying the five dissimilarity measures. For LOM measures, the e and g parameters generating the optimal ARI were selected (introduced in detail in Section 4.3.2). The Mantel test was then applied to calculate the correlation between each pair of dissimilarity matrices by calculating the correlation between the two sets of n(n – 1)/2 matrix elements. The correlation was calculated multiple times, with random permutations on one of the two matrices' columns and rows, to estimate the significance of the Mantel test result.

### 3.5.2 Degree of Agreement

By applying the five dissimilarity measures, under two different sequence encoding schemes, we received 10 sets of crash categorization. To compare the results, we evaluated their degrees of agreement with a benchmark, the CRSS crash categorization (as discussed in Section 3.4 and illustrated in Figure 5). To measure the agreement, we employed the Adjusted Rand Index (ARI), Adjusted Mutual Information (AMI), and Fowlkes–Mallows Score (FMS) (Fowlkes and Mallows, 1983; Hubert and Arabie, 1985; Vinh et al., 2010). Those three measures do not work exactly the same, but all have the same purpose of measuring the degree of agreement between two groupings of data points. Readers interested in differences between clustering comparison measures should refer to (Wagner and Wagner, 2007; Romano et al., 2016) We hope to reach a reliable evaluation conclusion based on results from applying multiple different measures.

With a set of clustering results, Y, and the benchmark crash categorization, X, a contingency table could be written as:

|     | $Y_1$ | $Y_2$ | ... | $Y_s$ | Sum |
|-----|-------|-------|-----|-------|-----|
| $X_1$ | $n_{11}$ | $n_{12}$ | ... | $n_{1s}$ | $a_1$ |
| $X_2$ | $n_{21}$ | $n_{22}$ | ... | $n_{2s}$ | $a_2$ |
| ... | ... | ... | ... | ... | ... |
| $X_r$ | $n_{r1}$ | $n_{r2}$ | ... | $n_{rs}$ | $a_r$ |
| Sum | $b_1$ | $b_2$ | ... | $b_s$ | $\sum_{ij} n_{ij} = N$ |

where $n_{ij}$ = the number of sequences assigned to both groups $X_i$ and $Y_j$, with $1 \leq i \leq r$ and $1 \leq j \leq s$; $a_i = \sum_{j=1}^{s} n_{ij}$; and $b_j = \sum_{i=1}^{r} n_{ij}$.

The ARI is calculated as:

$$ARI(X,Y) = \frac{2(N_{00}N_{11} - N_{01}N_{10})}{(N_{00}+N_{01})(N_{01}+N_{11})+(N_{00}+N_{10})(N_{10}+N_{11})} \qquad [6]$$

Where $N_{00}$ = the number of pairs that are in different clusters in both X and Y; $N_{01}$ = the number of pairs that are in the same cluster in X but in different clusters in Y; $N_{10}$ = the number of pairs that are in different clusters in X but in the same cluster in Y; $N_{11}$ = the number of pairs that are in the same clusters in both X and Y. The ARI ranges from -1 to 1. An ARI of 0 means a random agreement and the two groupings can be treated as independent. An ARI of 1 means the two groupings are identical (Kang et al., 2020; Yeung and Ruzzo, 2001).

The AMI is calculated as:

$$AMI(X,Y) = \frac{I(X,Y) - E\{I(X,Y)\}}{\max\{H(X),H(Y)\} - E\{I(X,Y)\}} \qquad [7]$$

Where $I(X,Y) = \sum_{i=1}^{r} \sum_{j=1}^{s} \frac{n_{ij}}{N} \log \frac{n_{ij}/N}{a_i b_j/N^2}$; $H(X) = -\sum_{i=1}^{r} \frac{a_i}{N} \log \frac{a_i}{N}$; $H(Y) = -\sum_{j=1}^{s} \frac{b_j}{N} \log \frac{b_j}{N}$. The AMI ranges from 0 to 1, with 1 meaning two groupings are identical, and 0 meaning they are independent (Vinh et al., 2010).

The FMS is calculated as:

$$FMS\ (X,Y) = \sqrt{\frac{N_{11}}{N_{11}+N_{10}} \cdot \frac{N_{11}}{N_{11}+N_{01}}} \qquad [8]$$

Where $N_{11}$, $N_{10}$, and $N_{01}$ are defined as in the ARI equation. Similar to the AMI, the FMS ranges from 0 to 1, with a higher score meaning a higher similarity between two groupings.

*3.6  R Packages*

Sequence analysis in this paper was carried out using R (R Core Team, 2019). All dissimilarity measures listed in Table 4 were available in the R sequence analysis package, "TraMineR", which was used as the primary tool for crash sequence analysis in this paper (Gabadinho et al., 2011). Other major R packages used in the analysis include "ade4" (for Mantel test), "WeightedCluster" (for weighted k-medoids clustering), , "aricode" (for ARI and AMI calculation), and "dendextend" (for FMS calculation) (Studer, 2013; Galili, 2015; Thioulouse et al., 2018; Chiquet et al., 2020).

## 4  Results and Discussion

*4.1  Crash Sequence Patterns*

Clustering was carried out using weighted k-medoids method on a sample of 2,676 interstate single-vehicle crashes. Five dissimilarity measures and two encoding schemes were employed, resulting in 10 sets of sequence clusters. To show some details of the results without listing all sets of clusters, we present an example of sequence clusters generated by employing the OMlev dissimilarity measure and both the original and new encoding schemes in Table 5. For each cluster, a description of the representative sequence is provided. A representative sequence is the dominant sequence in a cluster, making up the largest proportion, not all the sequences. Representative sequences were extracted by sorting sequences in each cluster. The 5 largest clusters under both the original and new encoding schemes are similar, represented by the following sequences:

- Vehicle was moving straight, then ran off road on the left side
- Vehicle was moving straight, then hit an animal, pedestrian, or bicyclist in its lane
- Vehicle was moving straight, then ran off the road on the right side
- Vehicle was negotiating a curve while speeding, then ran off the road on the left side
- Vehicle was moving straight while speeding, then ran off the road on the left side

**Table 5 Sequence clustering results with the OMlev measure**

**(a) Original Encoding**

| Cluster # | % | Representative Sequence Description | | |
|---|---|---|---|---|
| | | Pre-crash Events | Avoidance Maneuver | Collision Events |
| 1 | 14 | Moving Straight-RORL | Avoidance Unknown | Hit Fixed Object and/or Rollover |
| 2 | 14 | Moving Straight-Animal/Ped/Bike in Lane | Avoidance Unknown | Hit Animal/Ped/Bike |
| 3 | 11 | Moving Straight-RORR | Avoidance Unknown | Hit Fixed Object and/or Rollover |
| 4 | 9 | Negotiating Curve-Speeding-RORL | Avoidance Unknown | Hit Fixed Object and/or Rollover |
| 5 | 8 | Moving Straight-Speeding-RORL | Avoidance Unknown | Hit Fixed Object and/or Rollover |
| 6 | 6 | Negotiating Curve-RORL | Avoidance Unknown | Hit Fixed Object and/or Rollover |
| 7 | 5 | Moving Straight-RORL | No Avoidance | Hit Fixed Object and/or Rollover |
| 8 | 5 | Negotiating Curve-RORR | Avoidance Unknown | Hit Fixed Object and/or Rollover |
| 9 | 5 | Moving Straight-Object in Lane | Avoidance Unknown | Hit Object |
| 10 | 4 | Moving Straight-RORR | Steering R | Hit Fixed Object and/or Rollover |
| 11 | 4 | Moving Straight-Speeding-RORR | Avoidance Unknown | Hit Fixed Object and/or Rollover |
| 12 | 4 | Moving Straight-Other in Lane-ROR | Steering L | Hit Fixed Object and/or Rollover |
| 13 | 3 | Moving Straight-Poor Surface-ROR | Avoidance Unknown | Hit Fixed Object and/or Rollover |
| 14 | 3 | Moving Straight-Tire Issue-ROR | Avoidance Unknown | Hit Fixed Object and/or Rollover |
| 15 | 2 | Moving Straight-RORR | No Avoidance | Hit Fixed Object and/or Rollover |

Note: % of weighted sample total of 385,484 crashes.
   RORR = run-off-road right; RORL = run-off-road left; ROR = run-off-road (right or left).
   Steering R = steering right; Steering L = steering left.

**(b) New Encoding**

| Cluster # | % | Representative Sequence Description | | |
|---|---|---|---|---|
| | | Pre-crash Events | Avoidance Maneuver | Collision Events |
| 1 | 15 | Moving Straight-RORL | Avoidance N | Hit Fixed Object and/or Rollover |
| 2 | 13 | Moving Straight-Animal/Ped/Bike in Lane | Avoidance N | Hit Animal/Ped/Bike |
| 3 | 13 | Moving Straight-RORR | Avoidance N | Hit Fixed Object and/or Rollover |
| 4 | 9 | Negotiating Curve-Speeding-ROR | Avoidance N | Hit Fixed Object and/or Rollover |
| 5 | 8 | Moving Straight-Speeding-RORL | Avoidance N | Hit Fixed Object and/or Rollover |
| 6 | 7 | Negotiating Curve-RORL | Avoidance N | Hit Fixed Object and/or Rollover |
| 7 | 5 | Moving Straight-Speeding-Other Vehicle in Lane-RORL | Steering L | Hit Fixed Object and/or Rollover |
| 8 | 5 | Negotiating Curve-RORR | Avoidance N | Hit Fixed Object and/or Rollover |
| 9 | 5 | Moving Straight-Object in Lane | Avoidance N | Hit Object |
| 10 | 4 | Moving Straight-Other Vehicle in Lane | Steering R | Hit Fixed Object and/or Rollover |
| 11 | 4 | Moving Straight-Speeding-RORR | Avoidance N | Hit Fixed Object and/or Rollover |
| 12 | 4 | Moving Straight-Equip Failure-ROR | Avoidance N | Hit Fixed Object and/or Rollover |
| 13 | 3 | Moving Straight-Control Loss Other-ROR | Avoidance N | Hit Fixed Object and/or Rollover |
| 14 | 3 | Moving Straight-Control Loss-ROR | Avoidance N | Hit Fixed Object and/or Rollover |
| 15 | 2 | Moving Straight-Other/Unknown Event | Avoidance N | Hit Fixed Object and/or Rollover |

Note: % of weighted sample total of 385,484 crashes; Avoidance N = avoidance unknown / no avoidance.

Specific sequences assigned to each cluster may be different, as illustrated by the alluvial diagram in Figure 7. The diagram shows how the crashes were characterized by 1) the CRSS crash categorization, 2) sequence clustering with the original encoding scheme, and 3) sequence clustering with the new encoding scheme. Crashes are illustrated as alluvia and color-coded based on the CRSS crash configuration, the "flows" of the alluvia show how the crashes were re-grouped as characterization method switches. The categories or clusters are shown as stacks of boxes, with their heights indicating the proportion. Although the sequence clustering results under the original and new encoding schemes are similar, we can still observe regrouping of crashes and changes in the size of clusters moving from the original to the new encoding scheme.

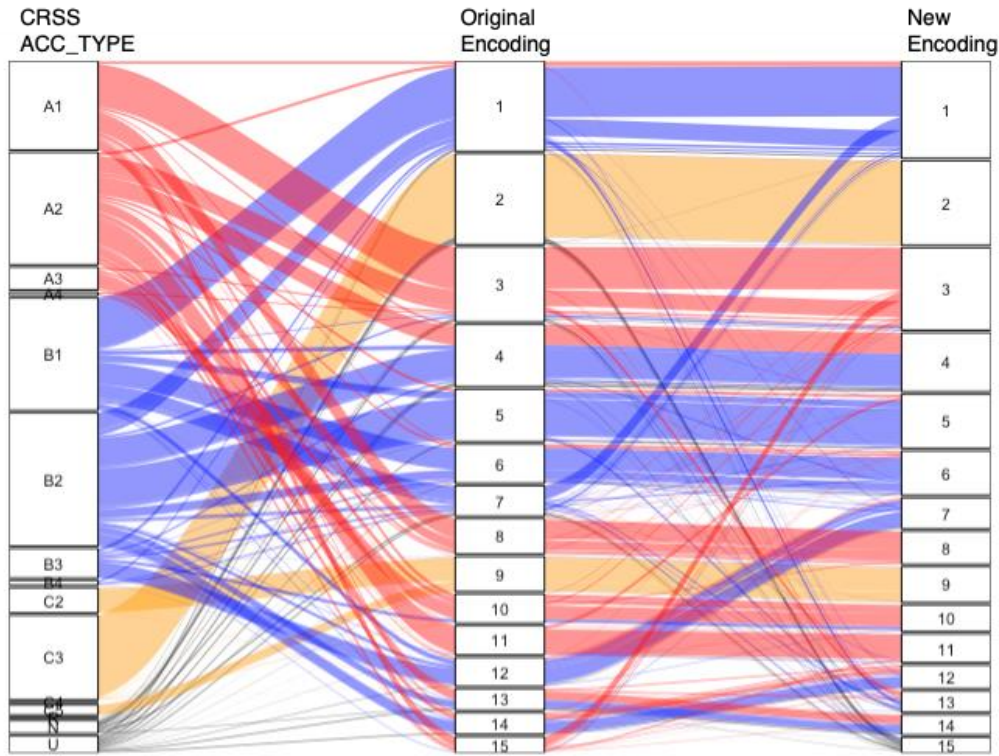

Note: Labels in boxes indicate the categories/clusters.

**Figure 7  Alluvial diagram of clustering results with the OMlev measure**

*4.2    Similarity Between Sequence Dissimilarity Measures*

Mantel test was applied to compare the crash sequence dissimilarity matrices generated from clustering with five dissimilarity measures. The Mantel test result ranges from -1 to 1, with a higher absolute value meaning a higher correlation between the two dissimilarity matrices. The results of Mantel correlation among the five OM dissimilarity measures, with the original and new encoding schemes, are shown in Table 6. The results are color coded with darker color meaning a higher correlation, and lighter color meaning a lower correlation. All results are statistically significant at the 0.01 level.

Mantel correlation matrices presented very similar patterns across the two encoding schemes. The results showed that OMlev, OMtr, and LOMtr generated highly positively correlated dissimilarity matrices. OMsf and LOMsf were highly correlated. LOMsf also had moderately high correlations with OMlev, OMtr, and LOMtr. Therefore, based on the Mantel correlations, the five dissimilarity measures could be categorized as the following two groups: Group 1: OMlev, OMtr, and LOMtr; Group 2: OMsf and LOMsf.

**Table 6  Mantel test results**

**(a) Original Encoding**

|       | OMlev | OMtr | OMsf | LOMtr | LOMsf |
|-------|-------|------|------|-------|-------|
| OMlev |       | 0.98 | 0.63 | 0.98  | 0.78  |
| OMtr  | 0.98  |      | 0.66 | 0.98  | 0.81  |
| OMsf  | 0.63  | 0.66 |      | 0.74  | 0.90  |
| LOMtr | 0.98  | 0.98 | 0.74 |       | 0.83  |
| LOMsf | 0.78  | 0.81 | 0.90 | 0.83  |       |

**(b) New Encoding**

|       | OMlev | OMtr | OMsf | LOMtr | LOMsf |
|-------|-------|------|------|-------|-------|
| OMlev |       | 0.98 | 0.64 | 0.97  | 0.77  |
| OMtr  | 0.98  |      | 0.65 | 0.98  | 0.78  |
| OMsf  | 0.64  | 0.65 |      | 0.73  | 0.91  |
| LOMtr | 0.97  | 0.98 | 0.73 |       | 0.80  |
| LOMsf | 0.77  | 0.78 | 0.91 | 0.80  |       |

*4.3   Comparison of Sequence Clustering Performance*

Sequence clustering was conducted on 10 dissimilarity matrices generated by the five dissimilarity measures and two encoding schemes, using the weighted k-medoids method. A 15-cluster characterization of crash sequences was obtained for each combination of dissimilarity measure and encoding scheme. For each set of clustering results, an ARI score was calculated to measure its agreement with the CRSS crash characterization. The ARI scores were then compared to evaluate the sequence clustering performance of the 10 dissimilarity and encoding scheme combinations.

For LOM measures (LOMtr and LOMsf) parameters e and g are required to compute indel costs. A sensitivity analysis was conducted to identify changes in ARI with changes in indel cost parameters and find the parameter values for the optimal ARIs. The results of sensitivity analysis are plotted as shown in Figure 8. Indel cost parameter e (the weight on the maximum substitution cost) values ranging from 0 to 0.4 with an increment of 0.01 were tested. The corresponding parameter g (the weight on the average of substitution costs between inserted element and its adjacent elements) values were set based on the relationship of $g = 1 - 2e$. Based on the plots, we found that in general, a very small e value (e.g., around 0.1) would lead to a good ARI value. Nevertheless, a sensitivity test would be helpful to identify the optimal parameter settings for specific combination of sequence encoding scheme and dissimilarity measure.

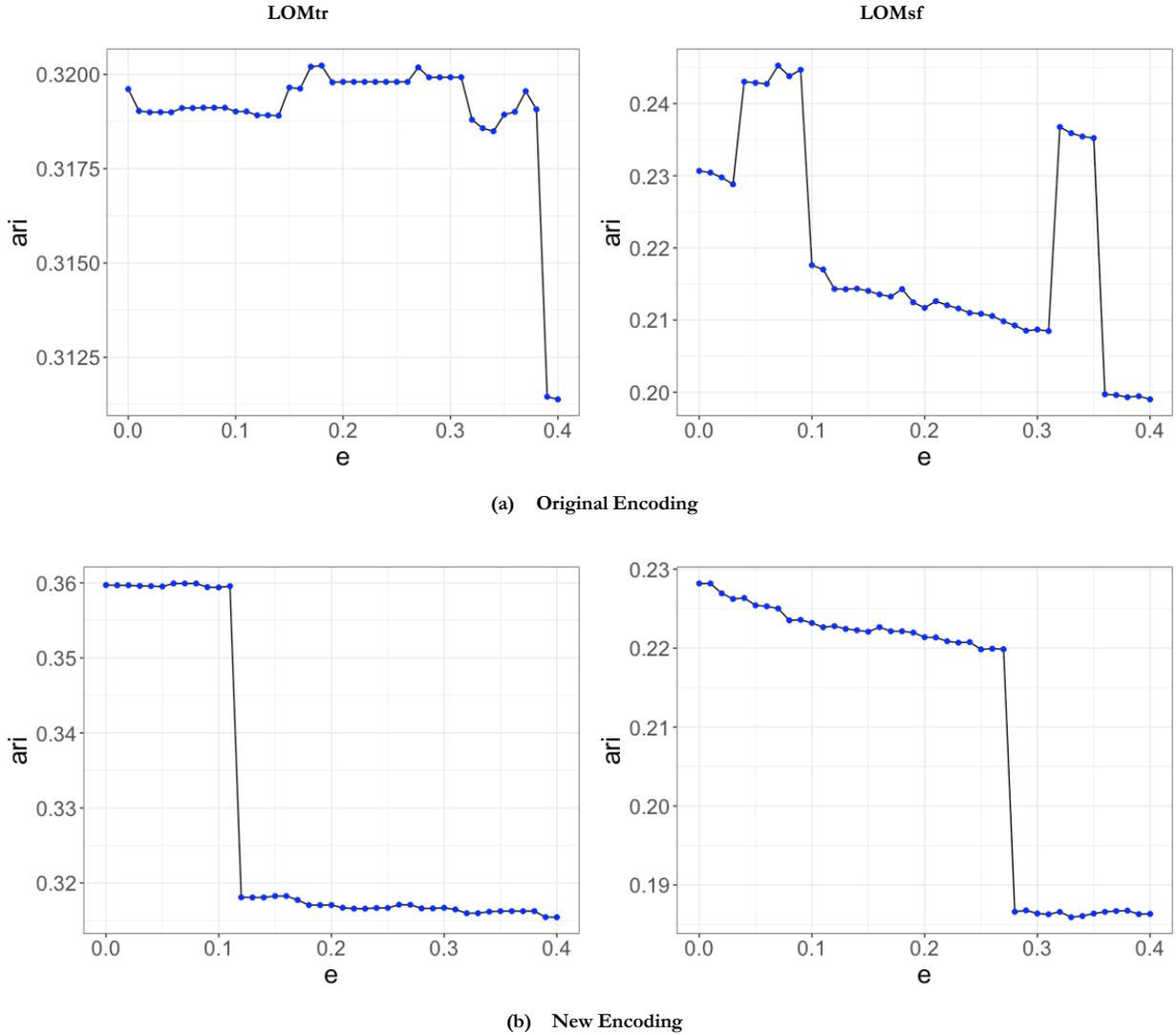

**Figure 8** ARI sensitivity to LOM parameter e

Comparing the 10 combinations of sequence dissimilarity measures and encoding schemes, the LOMtr measure obtained the largest ARI (0.319 with OE and 0.359 with NE) and the largest FMS (0.405 with OE and 0.441 with NE) under both encoding schemes, as shown in Table 7. However, the AMI results show that OMtr performed best with OE (0.411) and the OMlev measure performed best with NE (0.433). Overall, the Group 1 measures performed similarly well, and all had higher agreement scores than the Group 2 measures. The nature of crash sequence of events and the sequence structure applied in this paper (PCRASH+SOE) may be the reason that the shared-future cost scheme did not perform as well as the others. The different results of AMI from those of ARI and FMS can be explained by the fact that AMI is an information theoretic based measure, while ARI and FMS are both pair counting based measures (Vinh et al., 2010). Comparing the index values across encoding schemes, we found that in general, the Group 1 measures scored higher with NE and the Group 2 measures scored higher with OE. The reason may be that the shared-future model does not work well when the types of events were consolidated from OE to NE, which worked better with the Group 1 measures. In other words, as we chose to naturally consider and process context in sequence encoding (i.e., from OE to NE), using a context-sensitive dissimilarity measure upon a consolidated encoding may not offer a better clustering performance. Therefore, applying multiple dissimilarity measures and comparing the results is recommended for any crash sequence analysis.

**Table 7  Degrees of agreement**

| Measure | Group | ARI | | AMI | | FMS | |
|---|---|---|---|---|---|---|---|
| | | OE | NE | OE | NE | OE | NE |
| OMlev | 1 | 0.313 | 0.355 | 0.407 | **0.433** | 0.398 | 0.437 |
| OMtr | 1 | 0.313 | 0.316 | **0.411** | 0.407 | 0.398 | 0.402 |
| OMsf | 2 | 0.193 | 0.171 | 0.295 | 0.285 | 0.289 | 0.266 |
| LOMtr* | 1 | **0.320** | **0.359** | 0.409 | 0.425 | **0.405** | **0.441** |
| LOMsf* | 2 | 0.245 | 0.228 | 0.356 | 0.329 | 0.337 | 0.322 |

Note: * Showing largest values observed in sensitivity analysis.
OE = Original Encoding; NE = New Encoding.

In the study of single-vehicle roadside departure crash sequences by Wu et al., a Cohen's Kappa statistic (ranges from -1 to 1) was used to measure the agreement between clustering results and the original crash categorization in FARS, and a 0.28 Kappa value was deemed satisfactory (Wu et al., 2016). The ARI is equivalent to Cohen's Kappa (Warrens, 2008). Therefore, an ARI about 0.28 or larger was considered satisfactory in this paper. Regardless of the absolute values of ARI and the other two indices, for the purpose of comparison, emphasis was given to the relativity, and a larger index value means a better clustering performance.

Combining the findings from the Mantel tests and the clustering performance evaluation, we found that highly correlated dissimilarity measures performed similarly in clustering. In this interstate single-vehicle crash sequence analysis, Group 1 dissimilarity measures performed better and yielded larger ARIs than Group 2 measures. Comparing with LOMtr, the OMlev measure performed similarly well and needed less effort (i.e., lower computational cost) to generate a sequence dissimilarity matrix.

## 5    Conclusions

Crash sequence analysis supports crash characterization, which has importance in enhancing understanding of crashes and identifying safety countermeasures. Sequence analysis has been adapted recently to traffic crash study and shown to be effective in characterizing crashes and providing new insights. However, there is a need for a comparison study to evaluate the performance of various sequence encoding schemes and dissimilarity measures to help select the most appropriate techniques for crash sequence analysis.

In this paper, the impact of encoding and dissimilarity measures on crash sequence analysis and characterization was evaluated using interstate single-vehicle crash sequence data from the NHTSA CRSS database. Five optimal matching based dissimilarity measures were employed to compare crash sequences encoded with two schemes that represent two levels of information abstraction. A total of 2,676 crash sequences were characterized following a procedure of encoding, comparison, and clustering. The clustering results were evaluated by comparing with a benchmark crash categorization provided by the CRSS database.

The evaluation results suggest that encoding schemes and dissimilarity measures affect sequence comparison and sequence clustering. The transition-rate-based, localized optimal matching (LOMtr) dissimilarity measure, combining with a more abstract, consolidated encoding scheme, yielded the highest agreement scores with the benchmark crash categorization. The results indicate that the selection of dissimilarity measures plays a significant role in determining the results of sequence clustering and crash characterization. Also, dissimilarity measures that consider the relationships between events and domain context tend to perform well in crash sequence clustering. Encoding is essentially a prelude or preliminary step for measuring the dissimilarity between sequences, as the consolidation of similar types of events naturally takes domain context into consideration.

Practically, the sequence analysis procedure introduced in this paper is applicable to various sources of crash sequence data. The range of dissimilarity measures compared in this paper could also serve as a general list of candidates for crash sequence analysis. Traffic safety researchers and practitioners can follow this procedure to tune the encoding and dissimilarity measures for an appropriate crash characterization, which would help discover crash patterns and identify countermeasures.

There are limitations that could be addressed in future studies. Considering that the study was carried out on a specific set of crash sequences, the clustering and evaluation results may not be generalizable. The same dissimilarity measures or encoding schemes may perform differently on different sets of data. The selection of benchmark may also affect evaluation results. Apart from using an existing crash categorization, clustering benchmarks can be obtained in multiple ways such as expert-opinion-based clustering and crash classifications using criteria other than sequences (e.g., most harmful events, manner of collision, and injury severity) (Hennig, 2016). Future studies will extend to a variety of different crash types and apply different evaluation benchmarks. Moreover, in this paper, the influence of factors such as socio-demographic characteristics, traffic conditions, and weather were not considered in the modeling of sequences and characterization of crashes. More variables can be incorporated in future work and implemented in data subsetting or sequence encoding steps.

**Funding**

This research was partially sponsored by the Safety Research using Simulation University Transportation Center (SAFER-SIM). SAFER-SIM is funded by a grant from the U.S. Department of Transportation's University Transportation Centers Program (69A3551747131). The work presented in this paper remains the responsibility of the authors.

# Appendix A
# Table A - 1  Encoding schemes
**PCRASH1**

| Original | New | CRSS Description |
|---|---|---|
| 11p0 | N | 0 No Driver Present/Unknown if Driver Present |
| 11p1 | ST | 1 Going Straight |
| 11p2 | B | 2 Decelerating in Road |
| 11p3 | A | 3 Accelerating in Road |
| 11p4 | A | 4 Starting in Road |
| 11p5 | S | 5 Stopped in Roadway |
| 11p6 | PA | 6 Passing or Overtaking Another Vehicle |
| 11p10 | R | 10 Turning Right |
| 11p11 | L | 11 Turning Left |
| 11p12 | U | 12 Making a U-turn |
| 11p13 | BU | 13 Backing Up (Other Than for Parking Position) |
| 11p14 | C | 14 Negotiating a Curve |
| 11p15 | E | 15 Changing Lanes |
| 11p16 | E | 16 Merging |
| 11p17 | CA | 17 Successful Corrective Action to a Previous Critical Event |
| 11p98 | N | 98 Other |

**PCRASH2**

| Original | New | CRSS Description |
|---|---|---|
|  |  | *Loss of Control:* |
| 12p1 | LCS | 1 Blow Out/Flat Tire |
| 12p3 | LCS | 3 Disabling Vehicle Failure (e.g., Wheel Fell Off) |
| 12p4 | LCM | 4 Non-Disabling Vehicle Problem (e.g., Hood Flew Up) |
| 12p5 | LCM | 5 Poor Road Conditions (Puddle, Pothole, Ice, etc.) |
| 12p6 | LCF | 6 Traveling Too Fast for Conditions |
| 12p8 | LCO | 8 Other Cause of Control Loss |
| 12p9 | LCO | 9 Unknown Cause of Control Loss |
|  |  | *This Vehicle Traveling:* |
| 12p10 | ELL | 10 Over the Lane Line on Left Side of Travel Lane |
| 12p11 | ERL | 11 Over the Lane Line on Right Side of Travel Lane |
| 12p12 | ELE | 12 Off the Edge of The Road on The Left Side |
| 12p13 | ERE | 13 Off the Edge of The Road on The Right Side |
| 12p14 | ED | 14 End Departure |
| 12p15 | L | 15 Turning Left |
| 12p16 | R | 16 Turning Right |

| Original | New | CRSS Description |
|---|---|---|
| 12p19 | N | 19 Unknown Travel Direction |
| 12p20 | BU | 20 Backing |
| 12p21 | U | 21 Making a U-Turn |
| | | *Other Vehicle in Lane:* |
| 12p50 | OIS | 50 Other Vehicle Stopped |
| 12p51 | OIS | 51 Traveling in Same Direction with Lower Steady Speed |
| 12p52 | OIS | 52 Traveling in Same Direction while Decelerating |
| 12p59 | OIN | 59 Unknown Travel Direction of The Other Motor Vehicle in Lane |
| | | *Other Vehicle Encroaching into Lane:* |
| 12p60 | OES | 60 From Adjacent Lane (Same Direction)-Over Left Lane Line |
| 12p61 | OES | 61 From Adjacent Lane (Same Direction)-Over Right Lane Line |
| 12p62 | OEO | 62 From Opposite Direction Over Left Lane Line |
| 12p63 | OEO | 63 From Opposite Direction Over Right Lane Line |
| 12p64 | OES | 64 From Parking Lane/Shoulder, Median/Crossover, Roadside |
| 12p66 | OET | 66 From Crossing Street, Across Path |
| 12p74 | OES | 74 From Entrance to Limited Access Highway |
| 12p78 | OEN | 78 Encroaching by Other Vehicle – Details Unknown |
| | | *Pedestrian in Lane:* |
| 12p80 | PII | 80 Pedestrian in Road |
| 12p81 | PIA | 81 Pedestrian Approaching Road |
| | | *Pedalcyclist in Lane:* |
| 12p83 | BII | 83 Pedalcyclist/Other Non-Motorist in Road |
| 12p85 | BIN | 85 Pedalcyclist Or Other Non-Motorist Unknown Location |
| | | *Animal in Lane:* |
| 12p87 | AII | 87 Animal in Road |
| 12p88 | AIA | 88 Animal Approaching Road |
| 12p89 | AIN | 89 Animal Unknown Location |
| | | *Object in Lane:* |
| 12p90 | OBI | 90 Object in Road |
| 12p91 | OBA | 91 Object Approaching Road |
| | | *Other/Unknown:* |
| 12p98 | N | 98 Other Critical Precrash Event |
| 12p99 | N | 99 Unknown |

**PCRASH3**

| Original | New | CRSS Description |
|---|---|---|
| 13p0 | N | 0 No Driver Present/Unknown if Driver Present |
| 13p1 | N | 1 No Avoidance Maneuver |
| 13p5 | RB | 5 Releasing Brakes |
| 13p6 | L | 6 Steering Left |
| 13p7 | R | 7 Steering Right |
| 13p8 | BL | 8 Braking and Steering Left |
| 13p9 | BR | 9 Braking and Steering Right |
| 13p10 | A | 10 Accelerated |
| 13p12 | AR | 12 Accelerating and Steering Right |
| 13p15 | B | 15 Braking and Unknown Steering Direction |
| 13p16 | B | 16 Braking |
| 13p98 | N | 98 Other Actions |
| 13p99 | N | 99 Unknown/Not Reported |

**SOE**

| Original | New | CRSS Description |
|---|---|---|
| 1v1 | RLO | 1 Rollover/Overturn |
| 1v2 | NCH | 2 Fire/Explosion |
| 1v3 | NCH | 3 Immersion or Partial Immersion |
| 1v5 | NCH | 5 Fell/Jumped from Vehicle |
| 1v7 | NCH | 7 Other Noncollision |
| 1v8 | XP | 8 Pedestrian |
| 1v9 | XB | 9 Pedalcyclist |
| 1v11 | XA | 11 Live Animal |
| 1v15 | XP | 15 Non-Motorist on Personal Conveyance |
| 1v16 | NCH | 16 Thrown or Falling Object |
| 1v17 | XFC | 17 Boulder |
| 1v18 | XO | 18 Other Object Not Fixed |
| 1v19 | XFA | 19 Building |
| 1v20 | XFA | 20 Impact Attenuator/Crash Cushion |
| 1v21 | XFC | 21 Bridge Pier or Support |
| 1v23 | XFB | 23 Bridge Rail (Includes Parapet) |
| 1v24 | XFA | 24 Guardrail Face |
| 1v25 | XFB | 25 Concrete Traffic Barrier |
| 1v26 | XFB | 26 Other Traffic Barrier |
| 1v30 | XFC | 30 Utility Pole/Light Support |
| 1v31 | XFC | 31 Post, Pole or Other Support |
| 1v32 | XFC | 32 Culvert |

| Original | New | CRSS Description |
|---|---|---|
| 1v33 | XFB | 33 Curb |
| 1v34 | XFA | 34 Ditch |
| 1v35 | XFA | 35 Embankment |
| 1v38 | XFA | 38 Fence |
| 1v39 | XFB | 39 Wall |
| 1v40 | XFC | 40 Fire Hydrant |
| 1v41 | XFA | 41 Shrubbery |
| 1v42 | XFC | 42 Tree (Standing Only) |
| 1v43 | XFB | 43 Other Fixed Object |
| 1v44 | NCH | 44 Pavement Surface Irregularity (Ruts, Potholes, Grates, etc.) |
| 1v46 | XFC | 46 Traffic Signal Support |
| 1v48 | XFA | 48 Snow Bank |
| 1v50 | XFC | 50 Bridge Overhead Structure |
| 1v52 | XFC | 52 Guardrail End |
| 1v53 | XFB | 53 Mail Box |
| 1v57 | XFA | 57 Cable Barrier |
| 1v58 | XFA | 58 Ground |
| 1v59 | XFB | 59 Traffic Sign Support |
| 1v61 | EF | 61 Equipment Failure (blown tire, brake failure, etc.) |
| 1v63 | ROR | 63 Ran Off Roadway-Right |
| 1v64 | ROL | 64 Ran Off Roadway-Left |
| 1v65 | CM | 65 Cross Median |
| 1v67 | AIR | 67 Vehicle Went Airborne |
| 1v68 | CM | 68 Cross Centerline |
| 1v69 | RE | 69 Re-entering Roadway |
| 1v71 | ED | 71 End Departure |
| 1v72 | NCH | 72 Cargo/Equipment Loss, Shift, or Damage (Harmful) |
| 1v73 | XO | 73 Object That Had Fallen from Motor Vehicle In-Transport |
| 1v79 | RO | 79 Ran off Roadway - Direction Unknown |
| 1v91 | XO | 91 Unknown Object Not Fixed |
| 1v93 | XFB | 93 Unknown Fixed Object |
| 1v99 | N | 99 Reported as Unknown |